%% file: artikel.tex
\documentclass[twocolumn,showpacs,amsmath,amssymb]{revtex4}

\usepackage{graphicx}% Include figure files
\usepackage{dcolumn}% Align table columns on decimal point
\usepackage{bm}% bold math

\begin{document}

\title{Strain induced correlation gaps in carbon nanotubes}

\author{T. A. Gloor}
\email{t.gloor@epfl.ch}	
\author{F. Mila}%
\affiliation{%
Institut de th\'eorie des ph\'enom\`enes physiques\\
\'Ecole Polytechnique F\'ed\'erale de Lausanne\\
BSP, CH-1015 Lausanne, Switzerland
}%

\date{\today}

\begin{abstract}
We calculate the change in the correlation gap of armchair carbon
nanotubes with uniaxial elastic strain. We predict that such a stretching
will enlarge the correlation gap for all carbon nanotubes by a change
that could be as large as several $\mathrm{meV}$ per percent of
applied strain, in contrast 
with pure band structure calculations where no change 
for armchair carbon nanotubes is predicted. The correlation effects
are considered within  a self--consistent Hartree--Fock approximation
to the Hubbard model with on--site repulsion only.  
\end{abstract}

 %PACS, the Physics and Astronomy
\maketitle

Carbon nanotubes (CNT) have many extraordinary electronic and
mechanical properties \cite{NT01}. In particular, band theory predicts
that CNT are either metallic or semiconducting depending on chirality,
i.e. in which direction a graphite monolayer is ``rolled up'' into a
cylinder forming the tube \cite{NT98}. Probing CNT with a scanning
tunneling microscope, this metallic or semiconducting behaviour could
be tested experimentally \cite{Wildoer98, Odom98}. In addition, CNT
can sustain large mechanical strains and can be deformed elastically
up to bendings of order $19^\circ$ which corresponds to a strain along the
tube of $5.5\%$ \cite{Tombler00}. Experiments and numerical
calculations indicate a large Young modulus of order $1 \mathrm{TPa}$
\cite{Salvetat99}. The interaction of mechanical and electronic
properties has been studied at room temperature in two experiments
\cite{Tombler00,Minot03} where it has been shown that uniaxial stress
can change dramatically 
the electronic structure of CNT. In these experiments CNT have been
suspended between two metal electrodes on $SiO_2/Si$ substrates. To
apply uniaxial stress along the tube the tip of an atomic force
microscope (AFM) was used. The tip was lowered to push at the center
of the CNT. The AFM allows to measure simultaneously  the deflection
and the conductance of the CNT. The strain can be defined as
$\sigma=\left[\sqrt{4\delta^2 + l^2}-l\right]/l$ where $l$ is the
suspended length of the tube and $\delta$ is its vertical
deflection. In the first experiment \cite{Tombler00} it was shown that
changing the strain from $0\%$ to $3.2\%$ let the conductance of a
metallic CNT drop by two orders of magnitude. Both the mechanical and
the electronic properties were observed to be completely
reversible. More recently it was demonstrated \cite{Minot03} that not only
metallic CNT become less conducting when applying stress but also
inversely that some samples modified their behaviour from
semiconducting to metallic. These experiments show convincingly that
uniaxial stress applied to CNT changes their electronic properties.
\\
Theoretically, the effect of strain on the electronic properties of
CNT has been studied in band structure calculations, either
analytically, using a tight--binding approach, or numerically by
density functional theory
\cite{Yang00,Heyd97,Kane97,Buongiorno99,Maiti02,Rochefort99}. In
particular it has been shown that, depending on chirality, uniaxial
stress can increase, decrease or not alter the band gap.\\
In this paper
we include the effect of electron--electron correlations in these
calculations. We compute in detail  the 
gap as a function of applied strain and we compare our results to the
one--electron band structure calculations in the literature. The
calculations are carried out within a Hubbard 
tight--binding model using the self--consistent Hartree--Fock (H--F)
approximation. It has been argued by the authors \cite{GloorMila03a}
that the charge gap of CNT at half-filling is well described within
this approximation.
\\

\begin{figure}
\input{honey_pg2.pstex_t}
\caption{\label{fig:honeycomb} A rectangular honeycomb lattice of a
$(4,2)$ CNT, i.e. $\bm{C_h} = 4\bm{a_1}+2\bm{a_2}$. The basis vectors
are chosen to 
be $\bm{a_1}=a/2(\sqrt{3},1) $ and
$\bm{a_2}=a/2(\sqrt{3},-1)$. $a=2.49\AA$ is the lattice constant for CNT. The
chirality vector $\bm{C_h}$ and the vector $\bm{T}$ define the
quasi--one--dimensional unit cell.  $\bm{C_h}$ describes the
circumference and  $\bm{T}$ is oriented parallel to the tube.}
\end{figure}
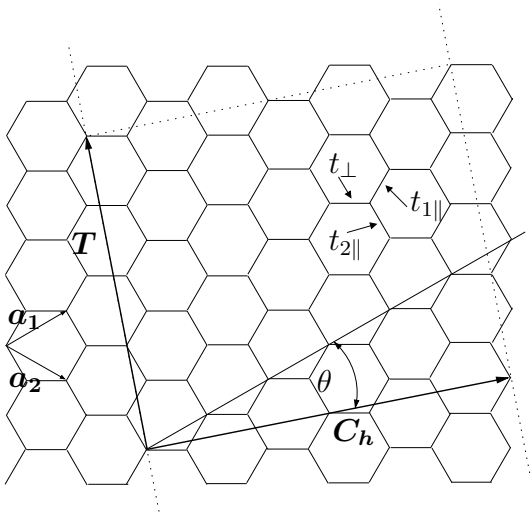

In this work we consider single--walled CNT at
half--filling. In a first approximation, single walled CNT can be
thought to be a rectangular 
graphite monolayer with the appropriate boundary conditions
(fig. \ref{fig:honeycomb}). 
They can be classified by their chirality
vector $\bm{C_h} = n\bm{a_1} + m\bm{a_2}$, where $\bm{a_1}$ and $\bm{a_2}$
are the basis vector of the honeycomb lattice, while $n$ and $m$ are integers
with $m\leq n$ \cite{NT98}. $\bm{C_h}$ determines into which direction the
graphene layer is rolled up. We model CNT with a Hubbard model with
nearest neighbour hopping between $\pi$ orbitals only, and on--site interaction at half--filling:
\begin{equation}
H = \sum_{\left\langle i,j\right\rangle \sigma}
\left(t_{ij}c^\dagger_{i\sigma} c_{j\sigma} +\mathrm{h.c.}\right) + U\sum_i n_{i\uparrow}n_{i\downarrow}
\label{eqn:hamiltonian}
\end{equation}
$\sigma$ is the spin index and $i,j$ sum over the sites of a
rectangular honeycomb lattice with periodic boundary
conditions. $c_{i\sigma}^\dagger$ ($c_{i\sigma}$) are the fermion
creation (annihilation) operators
and $n_{i\sigma} = c_{i\sigma}^\dagger c_{i\sigma}$. 
The hopping integrals $t_{ij}$ are restricted to nearest neighbors and 
in general they can have different values in every hopping direction,
say $t_\perp$, $t_{1\parallel}$ and $t_{2\parallel}$ (cf
fig. \ref{fig:honeycomb}). These three hopping amplitudes and the
one--site interaction strength $U$ are the four parameters entering
into the model. Choosing  some values for those parameters  we can
determine  the charge gap through a H--F calculation for a given 
$(n,m)$ CNT. All the details
of the H--F approximation can be found in reference
\cite{GloorMila03a} where also its validity has been discussed.\\
If there is no applied strain we set in our calculation all the hopping
amplitudes equal to a certain value $t_0$. This is of course only
an approximation. Forming a CNT from a graphite sheet, where this is true, we
change the hopping integrals due to effects of curvature. It has been
shown \cite{Kane97,Kleiner01} that there is a curvature induced gap
due to $\sigma$--$\pi$ hybridization for all CNT except armchair
CNT. This gap is at least of order $10$ 
$\mathrm{meV}$. Thus armchair CNT are the only CNT where band structure
calculations predict a metallic gapless behaviour. They correspond to
chiralities with $n=m$ or 
equivalently to a chiral angle $\theta=\pi/6$ (cf
fig. \ref{fig:honeycomb}). For armchair CNT the curvature induces only
two different hopping amplitudes, namely $t_\perp$ and
$t_\parallel\equiv t_{1\parallel}= t_{2\parallel}$. Due to this
symmetry, no gap is opened from the band point of view. However we will show that there will be a gap induced from electron--electron
correlation effects for all CNT, even if they are of armchair type. We
concentrate our discussion on armchair CNT. The same effects
are apparent in CNT of other chiralities but in that case these
effects are less visible as there is already a gap in the
one--electron band structure. \\
Following references \cite{Yang00,Heyd97} we use Harrison's
phenomenological law to relate the hopping parameter $t_0$ of the
undeformed CNT to the ones of the elastically deformed CNT 
\cite{Harrison} $t_i=t_0\left(r'_i/r_i\right)^2$ where $\bm{r'_{i}}$
and $\bm{r_i}$ respectively, are 
the bond vectors before and after the deformation and $i=\perp$,
$1_\parallel$, $2_\parallel$. Projecting these 
vectors along the directions of $\bm{T}$ and $\bm{C_h}$, we can write
for an elastic uniaxial strain along the tube: $r_{iT}=(1+\sigma)r'_{iT}$
and  $r_{iC_h}=(1-\nu\sigma)r'_{iC_h}$ where $\nu$ is the Poisson
ratio. $t_0$ can be estimated from ab initio calculations to be $2.4$ 
$\mathrm{eV}$ \cite{Mintmire92} and the Poisson ratio has been
computed numerically \cite{NT01} and 
measured experimentally for graphite to be about $\nu=0.2$
\cite{Graphite81}. It is difficult to get 
an estimate for the on--site repulsion $U$ of atomic carbon. In the
literature values between $5$ and $12$ $\mathrm{eV}$ are suggested
\cite{Jeckelmann95,Chakravarty92,Gunnarsson92}.\\

\begin{figure}
\includegraphics[width=5.7cm]{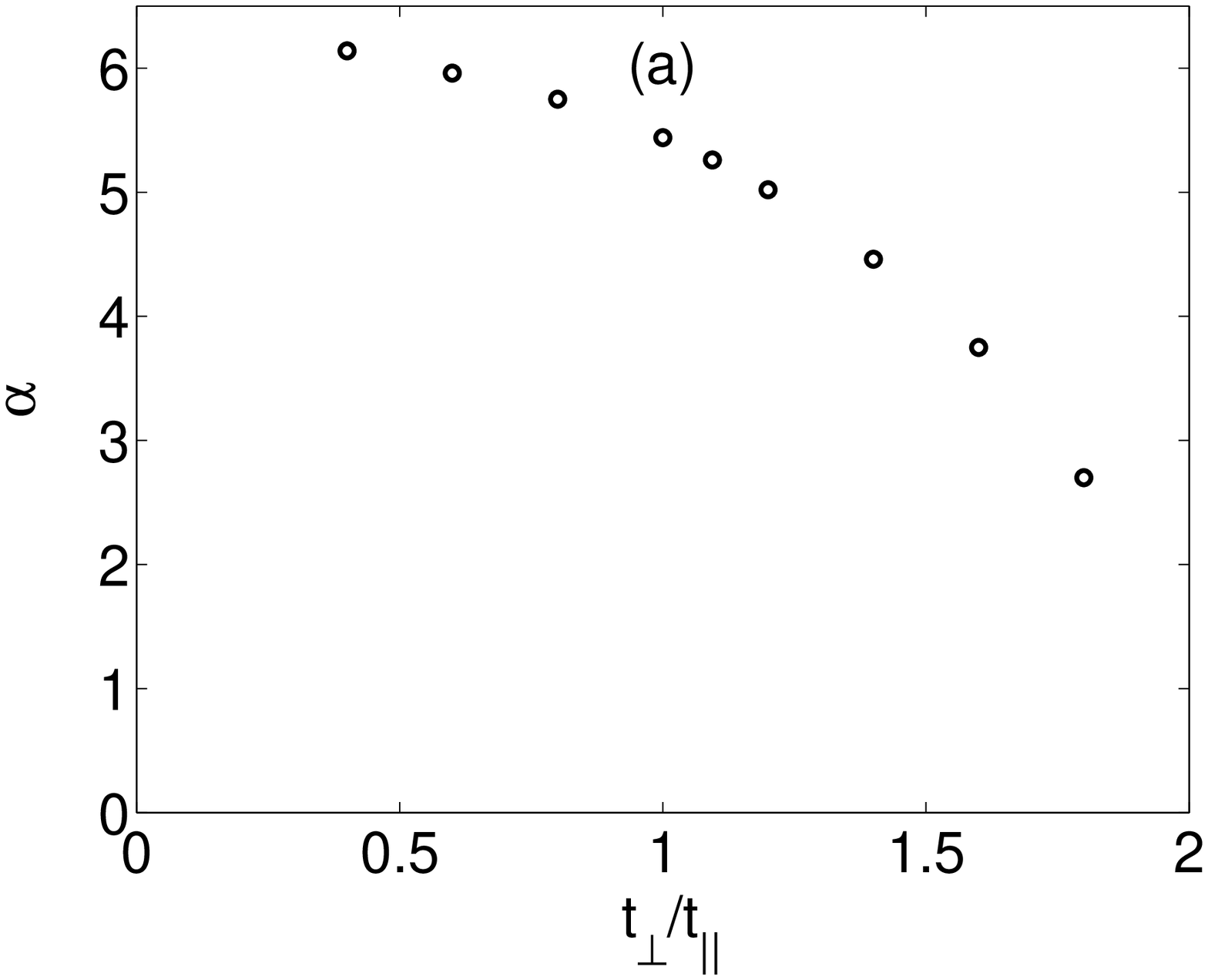}
\vspace{0.5cm}
\includegraphics[width=5.7cm]{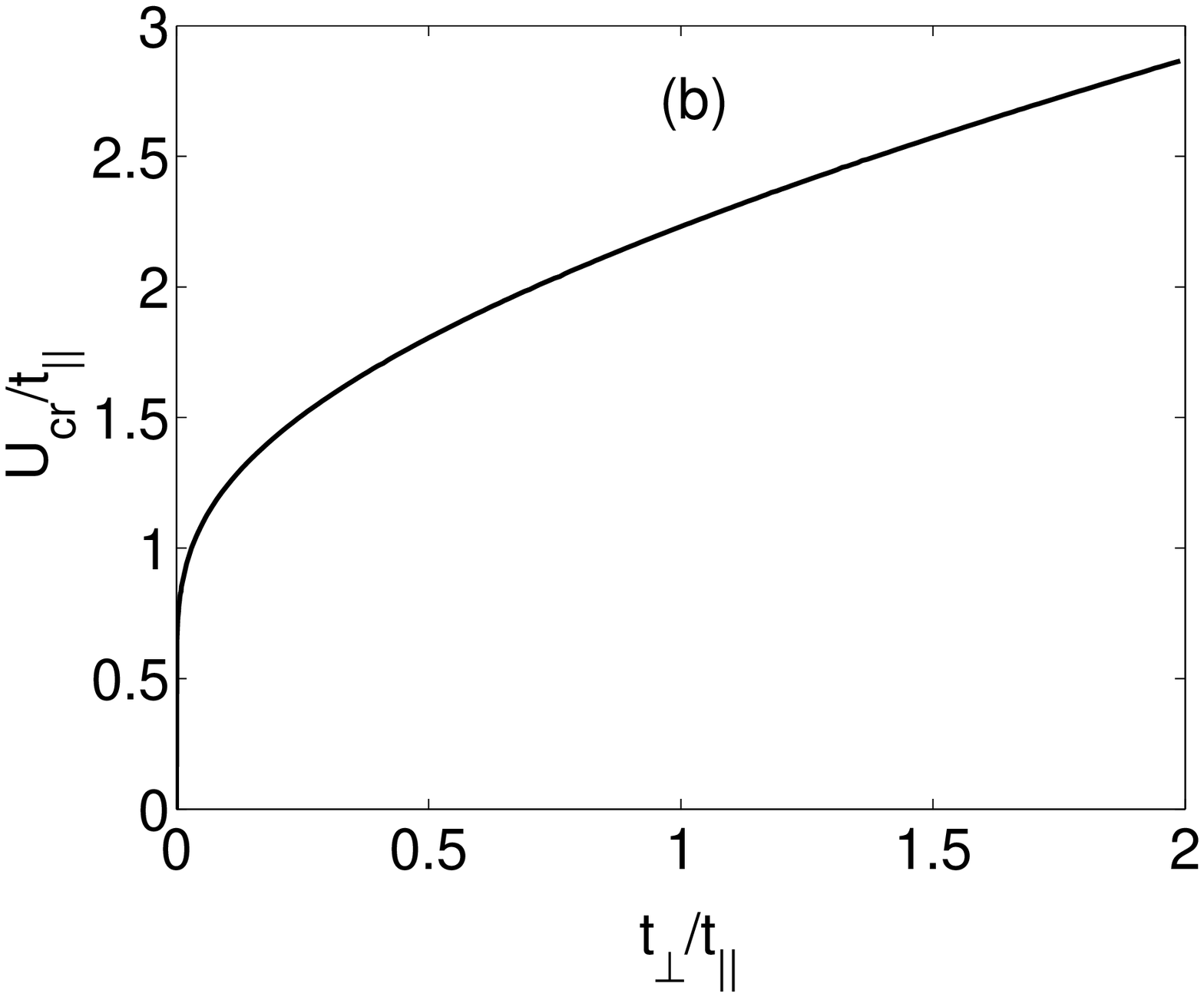}
\caption{\label{fig:HFparameters}Numerical results for the dependence
of the parameters $\alpha$ and $U_{cr}^{HF}$ in equation
(\ref{eqn:scalinglaw}) on $t_\perp/t_\parallel$:
(a) $\alpha$ is extracted from fitting the H--F results to the scaling
law (\ref{eqn:scalinglaw}). (b) $U_{cr}^{HF}$ is calculated by
evaluating $\chi_0 =
\int_{\epsilon<0}\mathrm{d}\epsilon\frac{\rho(\epsilon)}{2\left|\epsilon\right|}$,
the bare staggered static susceptibility, and where $\rho(\epsilon)$
is the tight--binding density of states of the honeycomb lattice. In
H--F theory we have $U_{cr}^{HF}=\chi_0^{-1}$.} 
\end{figure}

To study now the correlation effects in the situation described above,
i.e. stretching the tube by application of uniaxial strain, we rely as in
previous work 
\cite{GloorMila03a} on the H--F approximation. Based on the
observation that the H--F calculations reproduce correctly the
functional dependence of the energy gap in the one--dimensional
Hubbard model at half--filling, we believe that it gives also reliable
results for the gap of CNT.\\
We have shown in Ref. \cite{GloorMila03a} that the H--F calculations for the
charge gap of armchair CNT give the following scaling law
\begin{equation}
E_g/t_\parallel = c/n\, \exp\left\{-\alpha n (t_\parallel/U -
t_\parallel/U_{cr}^{HF})\right\}
\label{eqn:scalinglaw}
\end{equation}
with $c=1.01$, $\alpha = 5.44$ and $U_{cr}^{HF}=2.23t_0$, the critical value to
open a gap in the two--dimensional honeycomb lattice. Since graphite
is a semi--metal $U$ is expected
to be smaller, but close to this value, i.e. $U\lesssim U_{cr}^{HF}$. For
such values a correlation gap of order $10$ $\mathrm{meV}$ is present
for CNT of small diameter. We found
that the scaling law (\ref{eqn:scalinglaw}) is still valid when a
uniaxial stress $\sigma$ is applied, which produces a change of the
ratio $t_\perp/t_\parallel$ in the hamiltonian (\ref{eqn:hamiltonian})
from $1$ to larger values.  $t_\perp/t_\parallel$ and $\sigma$ are
related by Harrison's formula in a non--linear way. The prefactor $c$
depends only slightly on the change in  $t_\perp/t_\parallel$. The
variation of  $\alpha$ and 
$U_{cr}^{HF}$ on the applied strain is much more significant and it is
shown in figure \ref{fig:HFparameters}. We observe that $U_{cr}^{HF}$
increases with applied strain. From equation (\ref{eqn:scalinglaw}) we
can see that this would imply that the charge gap diminishes since a
$U$ far from the critical value disfavours a large gap. However the
simultaneous decrease of the parameter $\alpha$ overcomes this
tendency and when both effects are taken into account, the charge gap
increases approximately linearly with the strain with a slope that
depends on $U$.  
Note that $U_{cr}^{HF}>0$ indicates that there is a metal--insulator
transition at a finite value $U=U_{cr}^{HF}$. This is due to the fact
that the density of states in the honeycomb lattice is vanishing at
the two Fermi points as
$\rho(\epsilon)\propto\left|\epsilon\right|$. In the limit
$t_\perp/t_\parallel\rightarrow 0$ we get the one--dimensional
behaviour where any infinitesimal electron--electron interaction can
open a gap. The other limiting point is $t_\perp/t_\parallel=2$ where
the two Fermi points collapse into one and a band gap appears for
values $t_\perp/t_\parallel>2$ which makes the system insulating
already for $U=0$.\\

\begin{figure}
\includegraphics[width=5.7cm]{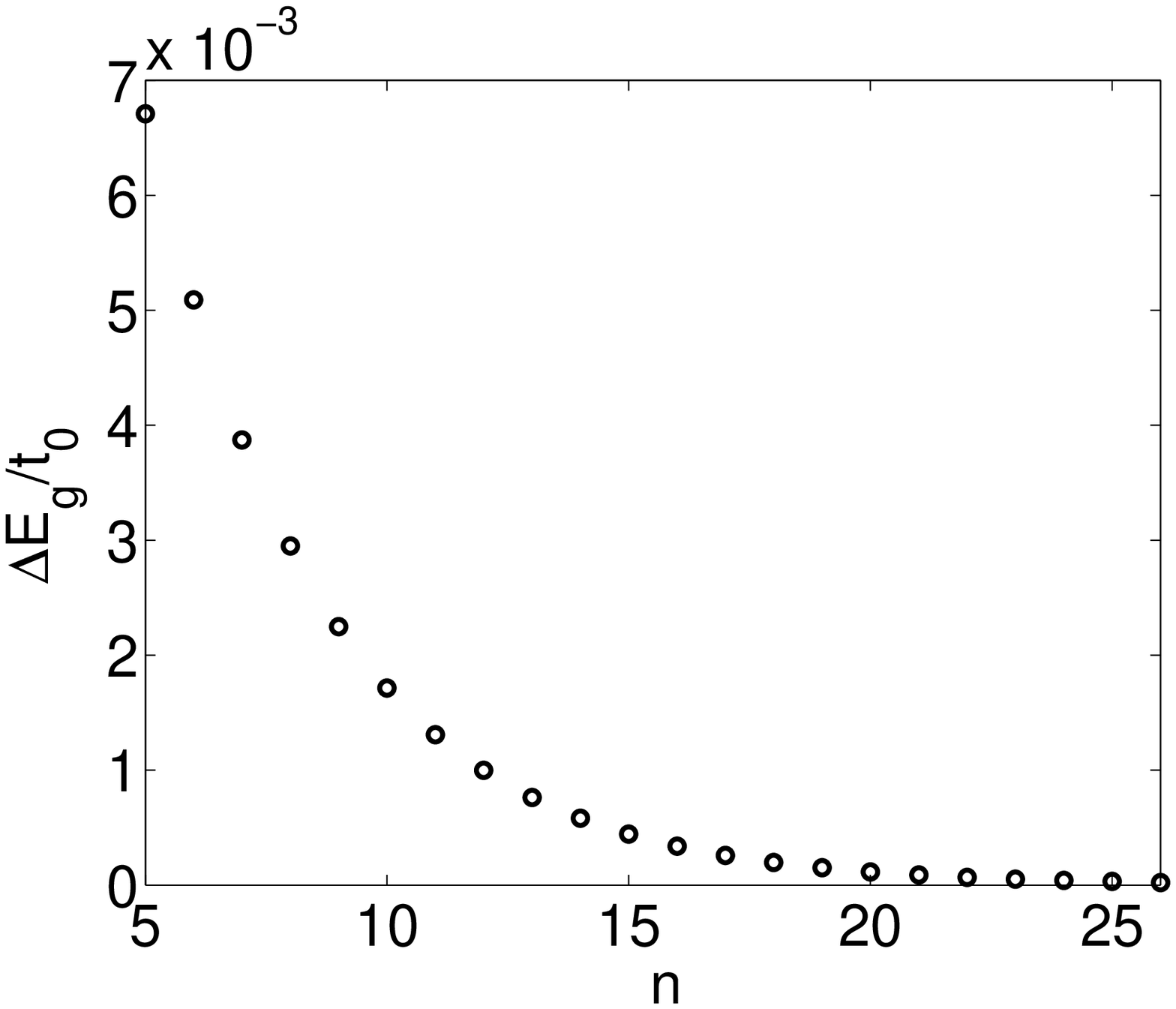}
\vspace{0.1cm}
\includegraphics[width=5.7cm]{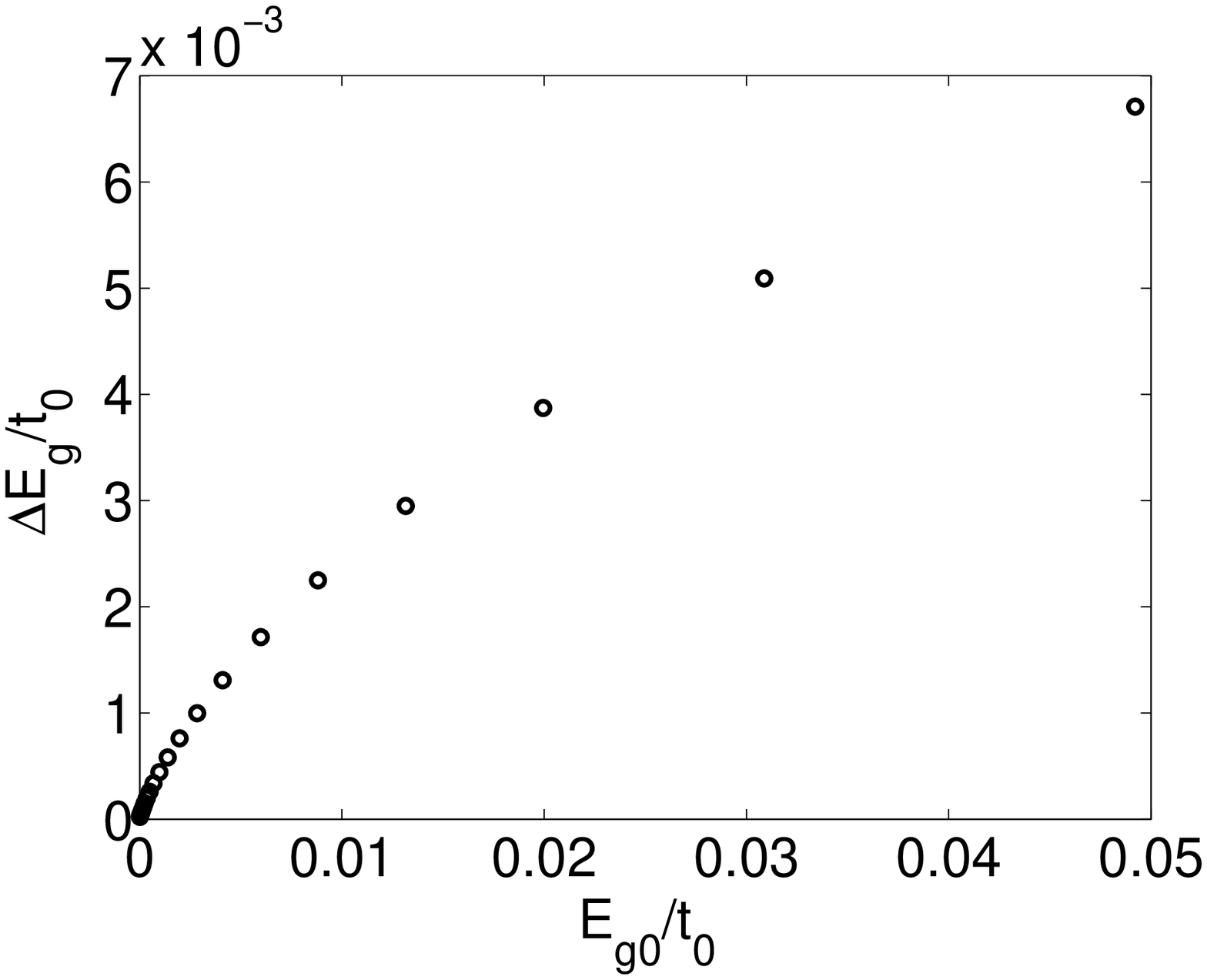}
\caption{\label{fig:ngap} The variation of the correlation gap for
different $(n,n)$ armchair CNT is plotted when a strain of $1\%$ is
applied ($U=2t_0$). The upper plot shows the variation of the gap
$E_g(\sigma=1\%) - E_g(\sigma=0)$ as a function of $n$. $n$ is
proportional to the tube diameter. The lower plot shows the same
variation as a function of the original correlation gap at zero applied strain
$E_{g0}\equiv E_g(\sigma=0)$. $E_{g0}$ is calculated for different
armchair CNT, $n=5,\dots , 26$.}
\end{figure}

To compare our results to experiments we plot the variation of the
gap for armchair CNT as a function of its size when a strain of $1\%$
is applied (cf figure \ref{fig:ngap}). The on--site interaction $U$ is
set to $2t_0$.  
As the hopping amplitude $t_0$ has a large value of $2.4$
$\mathrm{eV}$ neither the correlation gap at $\sigma=0$ of order $10$
$\mathrm{meV}$ nor its variation of order $\mathrm{meV}$ per percent
of applied strain can be neglected. As an example we look at the values
for a $(10,10)$ CNT. We can read off  from figure \ref{fig:ngap} that
the energy gap at zero strain is $14$ $\mathrm{meV}$ which corresponds
to a temperature of $160$ $\mathrm{K}$ and that a strain of $1\%$ induces  a
rise of $4$ $\mathrm{meV}$ in the gap.\\

How does this compare to the effect of strain to the
band structure? For small strains the following formula has been
derived from a tight--binding calculation \cite{Yang00}:
\begin{equation}
\label{eqn:band}
\frac{\mathrm{d}E_g}{\mathrm{d} \sigma} =
\mathrm{sgn}\left(2p+1\right)3t_0 \left(1+\nu\right) \cos 3\theta
\end{equation}
where $p\in\left\{-1,0,1\right\}$ is defined by the equation
$n-m=3q+p$ ($q$ is integer). This formula has been used to interpret
the experimental results in Ref. \cite{Minot03}. It follows from it
that the change of the band gap with applied
stress can be either positive or negative, depending on the value of
$q$, or in other words the chirality. The maximum variation is achieved for
zig--zag CNT (chiral angle zero) and is about $\pm85$
$\mathrm{meV}/\%$. The maximal variation of  $85$
$\mathrm{meV}/\%$ is one or even two orders of magnitude larger then
the variation of  $4$ $\mathrm{meV}/\%$ which we derived from
electron--electron correlation. However,
for armchair CNT ($\theta=\pi/6$) equation
(\ref{eqn:band}) and ab initio
calculations predict that no gap opens up with applied strain. Then
correlation effects are the only reason why one should have a
gap and this gap increases by applying uniaxial
strain at a rate of several $\mathrm{meV}/\%$. In our previous example
of the $(10,10)$ CNT, we have seen that one percent of strain can
change the correlation gap by about $30\%$ of its original value.\\

\begin{figure}
\includegraphics[width=5.8cm]{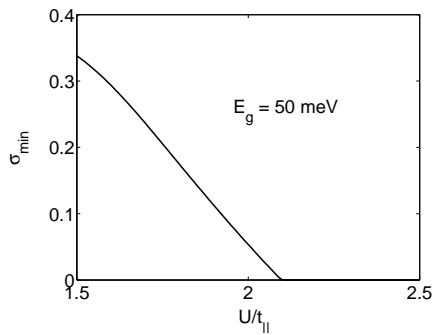}
\caption{\label{fig:50mev}  Minimal strain $\sigma_{min}$ necessary to
open a gap of $50\,\mathrm{meV}$ as a function of the on--site repulsion
$U$. The calculation was done for a $(10,10)$ armchair CNT.}
\end{figure}

To summarize, for semiconducting  CNT with large band gaps the
electronic structure at half filling is well described within band
theory and correlation effects give only small
corrections. But for CNT with a small gap (of order $\mathrm{meV}$)
correlation  
effects cannot be neglected. This is especially true for armchair CNT
where no band gap at all is predicted but they should develop a
measurable gap, induced from correlations alone, if sufficient pressure
is applied. This conclusion is illustrated in fig. \ref{fig:50mev}.
We plot the strain which is necessary to open a gap of  $E_g =
50\,\mathrm{meV}$ as a function of the on--site repulsion $U$ for a
$(10,10)$ armchair CNT. We see
that if $U$ is not too far from $U_{cr}$ this quite large gap would be
realisable experimentally and could be seen in low temperature
experiments.\\

This work was supported by the Swiss National Science Foundation and
by MaNEP.

\bibliography{nt}% Produces the bibliography via BibTeX.
\end{document}

%% file: honey_pg2.pstex_t
\begin{picture}(0,0)%
\includegraphics{honey_pg2.pstex}%
\end{picture}%
\setlength{\unitlength}{2362sp}%
\begingroup\makeatletter\ifx\SetFigFont\undefined%
\gdef\SetFigFont#1#2#3#4#5{%
  \reset@font\fontsize{#1}{#2pt}%
  \fontfamily{#3}\fontseries{#4}\fontshape{#5}%
  \selectfont}%
\fi\endgroup%
\begin{picture}(5533,5378)(540,-7293)
\put(589,-5300){\makebox(0,0)[lb]{\smash{\SetFigFont{12}{14.4}{\familydefault}{\mddefault}{\updefault}$\bm{a_1}$}}}
\put(3999,-6520){\makebox(0,0)[lb]{\smash{\SetFigFont{12}{14.4}{\rmdefault}{\mddefault}{\updefault}$\bm{C_h}$}}}
\put(3836,-6007){\makebox(0,0)[lb]{\smash{\SetFigFont{12}{14.4}{\rmdefault}{\mddefault}{\updefault}$\theta$}}}
\put(1261,-4558){\makebox(0,0)[lb]{\smash{\SetFigFont{12}{14.4}{\rmdefault}{\mddefault}{\updefault}$\bm{T}$}}}
\put(598,-5960){\makebox(0,0)[lb]{\smash{\SetFigFont{12}{14.4}{\familydefault}{\mddefault}{\updefault}$\bm{a_2}$}}}
\put(3961,-4516){\makebox(0,0)[lb]{\smash{\SetFigFont{12}{14.4}{\rmdefault}{\mddefault}{\updefault}$t_{2\parallel}$}}}
\put(3961,-3706){\makebox(0,0)[lb]{\smash{\SetFigFont{12}{14.4}{\rmdefault}{\mddefault}{\updefault}$t_\perp$}}}
\put(4816,-4111){\makebox(0,0)[lb]{\smash{\SetFigFont{12}{14.4}{\rmdefault}{\mddefault}{\updefault}$t_{1\parallel}$}}}
\end{picture}